\documentclass[10pt,sigconf,screen=true,anonymous=false,bookmarks=false,nonacm]{acmart}
\setcopyright{none}
\usepackage{enumitem}
\setlist{leftmargin=5.08mm}

\usepackage{lipsum}
\usepackage{graphicx}
\usepackage{amsmath}
\usepackage{footnote}
\usepackage{mathtools}
\usepackage{mathrsfs}     
\usepackage{comment}
\usepackage{soul}                                          
\usepackage[subrefformat=parens,labelformat=parens]{subfig}
\usepackage{bm,bbm}
\usepackage{multirow}
\usepackage{threeparttable,booktabs}
\usepackage{blkarray}
\usepackage{tikz}
\usetikzlibrary{positioning,calc,fit,decorations.pathmorphing,shapes.geometric,shapes.gates.logic.US,calc}
\usetikzlibrary{arrows,arrows.meta,decorations.markings,shapes,shapes.arrows}
\usetikzlibrary{patterns,snakes}
\usetikzlibrary{decorations,decorations.pathreplacing}
\usetikzlibrary{backgrounds}
\usepackage{tikz-3dplot}
\usepackage{balance}
\usepackage{courier}                                       
\usepackage{cleveref}                                      
\Crefformat{figure}{Fig.~#2#1#3}                          
\Crefname{subfigure}{Fig.}{Figs.}
\Crefname{figure}{Fig.}{Figs.}
\Crefname{equation}{Eq.}{Eqs.}
\usepackage[mathcal]{eucal}
\usepackage[]{algpseudocode}                               
\algrenewcommand\textproc{\texttt}
\makeatletter\let\float@addtolists\relax\makeatother
\usepackage{algorithm}
\usepackage{filecontents}                                  
\usepackage{pgfplots}
\usepackage{pgfplotstable}
\usepackage{pgf-pie}
\usepgfplotslibrary{groupplots}
\pgfplotsset{compat=newest}
\usepackage[figuresright]{rotating}
\usepackage{xcolor,colortbl}                               
\usepackage{adjustbox}                                     
\usepackage{makecell}
\usepackage{siunitx}
\usepackage{color}
\usepackage{diagbox}

\newcommand{\m}[1]{\boldsymbol{#1}}

\newcommand{\minisection}[1]{\vspace{.01in}\noindent{\textbf{#1}}.}

\theoremstyle{plain}

\theoremstyle{definition}

\algrenewcommand\textproc{\texttt}

\definecolor{CUHKorange}{RGB}{244,106,18} 
\definecolor{CUHKblue}{RGB}{0,111,190}    
\definecolor{CUHKgreen}{RGB}{0,127,128}   
\definecolor{CUHKred}{RGB}{228,46,36}     
\definecolor{CUHKyellow}{RGB}{198,148,34} 
\definecolor{CUHKdark}{RGB}{114,44,114}   
\definecolor{CUHKmiddle}{RGB}{144,44,144} 

\usepackage{tcolorbox}
\tcbuselibrary{skins,breakable}
    {\endtcolorbox}
    {\endtcolorbox}

\graphicspath{{./fig/figs/}{../}}

\usepackage{amsmath,amsfonts,bm}

\def\eqref#1{equation~\ref{#1}}

\def\1{\bm{1}}

\def\vv{{\bm{v}}}
\def\vw{{\bm{w}}}
\def\vx{{\bm{x}}}

\def\mV{{\bm{V}}}
\def\mW{{\bm{W}}}

\DeclareMathAlphabet{\mathsfit}{\encodingdefault}{\sfdefault}{m}{sl}
\SetMathAlphabet{\mathsfit}{bold}{\encodingdefault}{\sfdefault}{bx}{n}

\usepackage{siunitx}

\begin{document}
\date{}

\title{
    CSCO: A Backside-PDN-Aware \underline{C}lock–\underline{S}ignal \underline{C}o-\underline{O}ptimization Framework for Improved PPA
}

\author{
    Zixiao Wang$^1$,  \quad
    Leilei Jin$^{1,\dagger}$,   \quad
    Zhen Zhuang$^1$,  \quad 
    Rongmei Chen$^2$, \quad 
    Bei Yu$^{1,\dagger}$ \\
    $^1$ The Chinese University of Hong Kong \quad
    $^2$ Peking University
}

\begin{abstract}
    Backside power delivery networks (BSPDN) have emerged as a promising technology for advanced logic nodes to address IR-drop and PPA challenges. While BSPDN introduces additional routing resources on the backside, these resources are limited and must be carefully partitioned between clock and signal nets, creating a critical resource allocation tradeoff. Prior work either moves only the clock network or assumes a fixed clock tree and optimizes only signal nets, failing to explore the tradeoff space of backside resource allocation. Moreover, lacking frontside power-ground shielding, BSPDN introduces severe signal integrity (SI) degradation. We propose CSCO, a data-driven BSPDN-aware co-optimization framework that jointly allocates limited backside resources between clock and signal nets across frontside/backside layers. CSCO employs efficient search strategies to identify critical nets for backside routing without repeated evaluation, navigating the clock-signal allocation tradeoff to balance IR-drop, routing congestion, and PPA. The framework also leverages backside routing to mitigate coupling noise and crosstalk-induced SI issues. Experiments demonstrate improved WNS/TNS, frequency, and SI robustness without additional shielding overhead.
\end{abstract}

\maketitle
\hypersetup{pdfauthor={Zixiao Wang, Leilei Jin, Zhen Zhuang, Rongmei Chen, Bei Yu}}
\begingroup
\renewcommand{\thefootnote}{\fnsymbol{footnote}}
\footnotetext[2]{Corresponding authors.}
\endgroup
\thispagestyle{plain}
\pagestyle{plain}

\section{Introduction}
As technology scaling advances into the sub-3nm era, physical design faces increasingly severe back-end-of-line (BEOL) challenges including routing congestion, IR-drop, and reliability degradation~\cite{IEDM22,VLSI24,VLSI25,xie2025}. Metal resources traditionally allocated for signal routing and power delivery have become insufficient to sustain aggressive power, performance, and area (PPA) targets, making BEOL architecture and power delivery the dominant bottlenecks in system-level scaling. 
A promising solution has emerged through backside power delivery networks (BSPDN)~\cite{irdrop,LimVLSI25,A14-ppa}. By relocating the power grid to the wafer backside via nano-through-silicon-via (nTSV) connections, BSPDN alleviates IR-drop while liberating frontside metal layers for denser signal routing and higher operating frequencies as a critical enabler for next-generation design-technology co-optimization (DTCO) and 3D IC integration~\cite{pin3d,dac23,dtco,kim2024addressing,probe3,stco}.

\begin{figure}[t!]
  \centering
  \subfloat[FSPDN + frontside routing]{\includegraphics[width=0.493\linewidth]{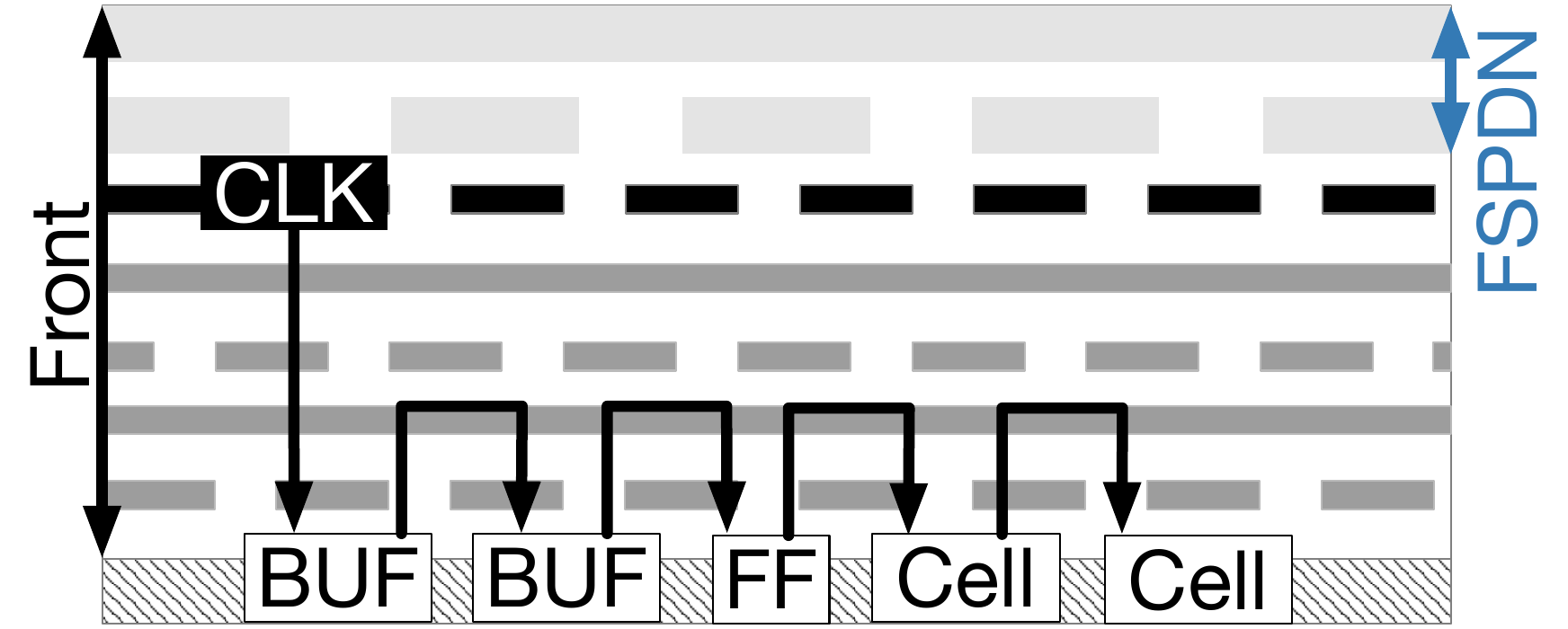}\label{fig:f1a}}
  \hfill
  \subfloat[BSPDN + backside clock only]{\includegraphics[width=0.493\linewidth]{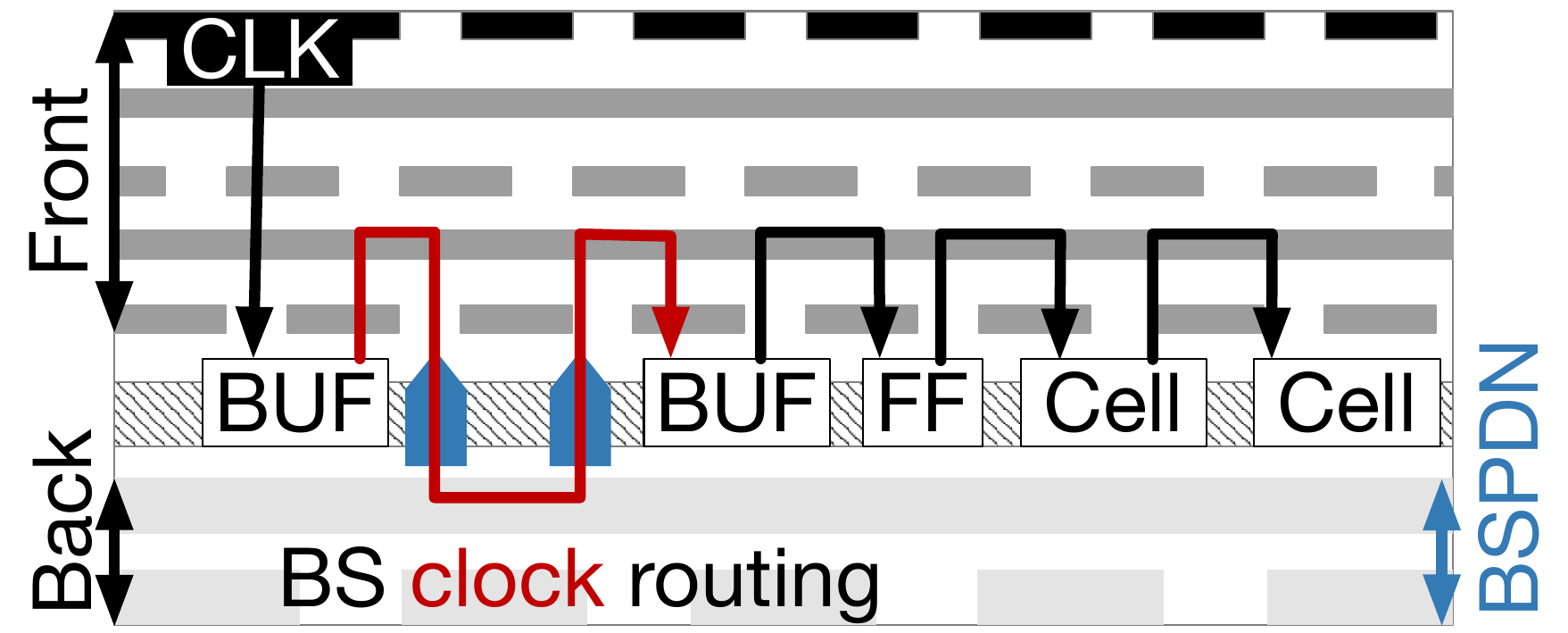}\label{fig:f1b}}\\
  \subfloat[BSPDN + backside signal only]{\includegraphics[width=0.493\linewidth]{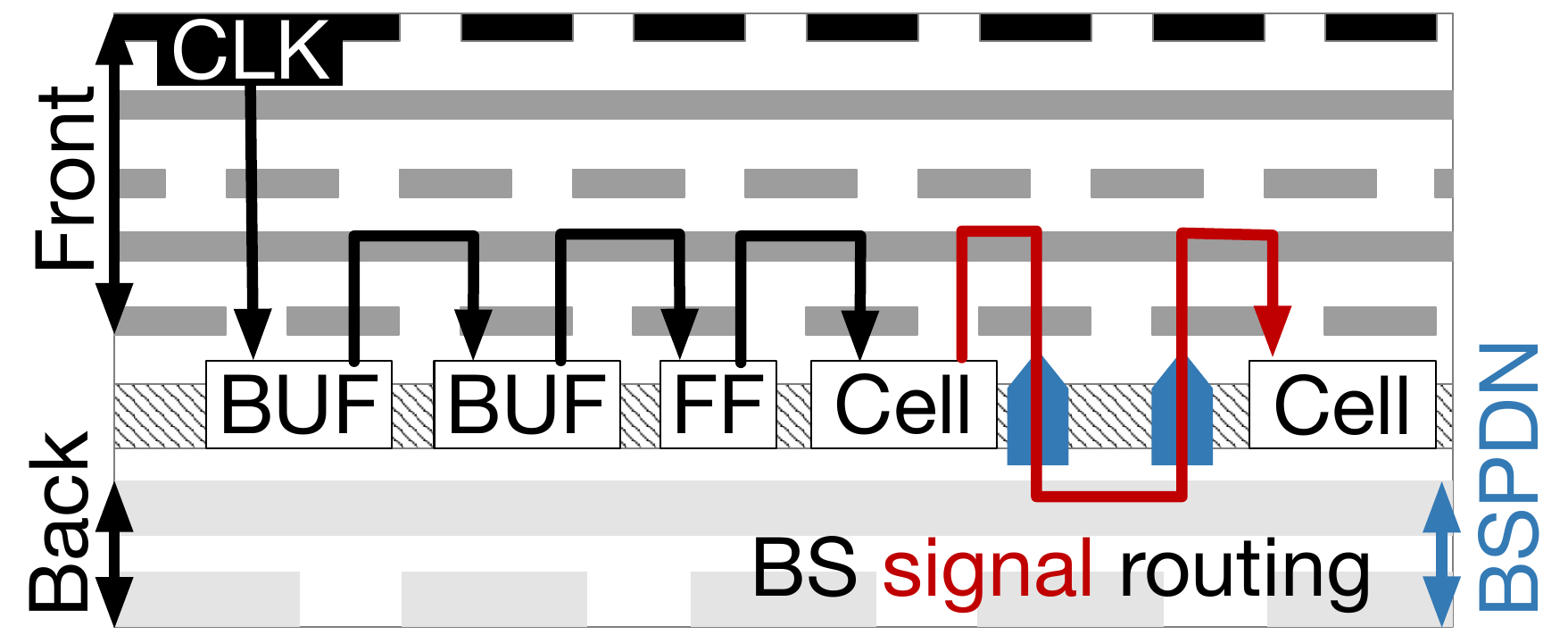}\label{fig:f1c}}
  \hfill
  \subfloat[BSPDN + double-side clock and signal routing (proposed)]{\includegraphics[width=0.493\linewidth]{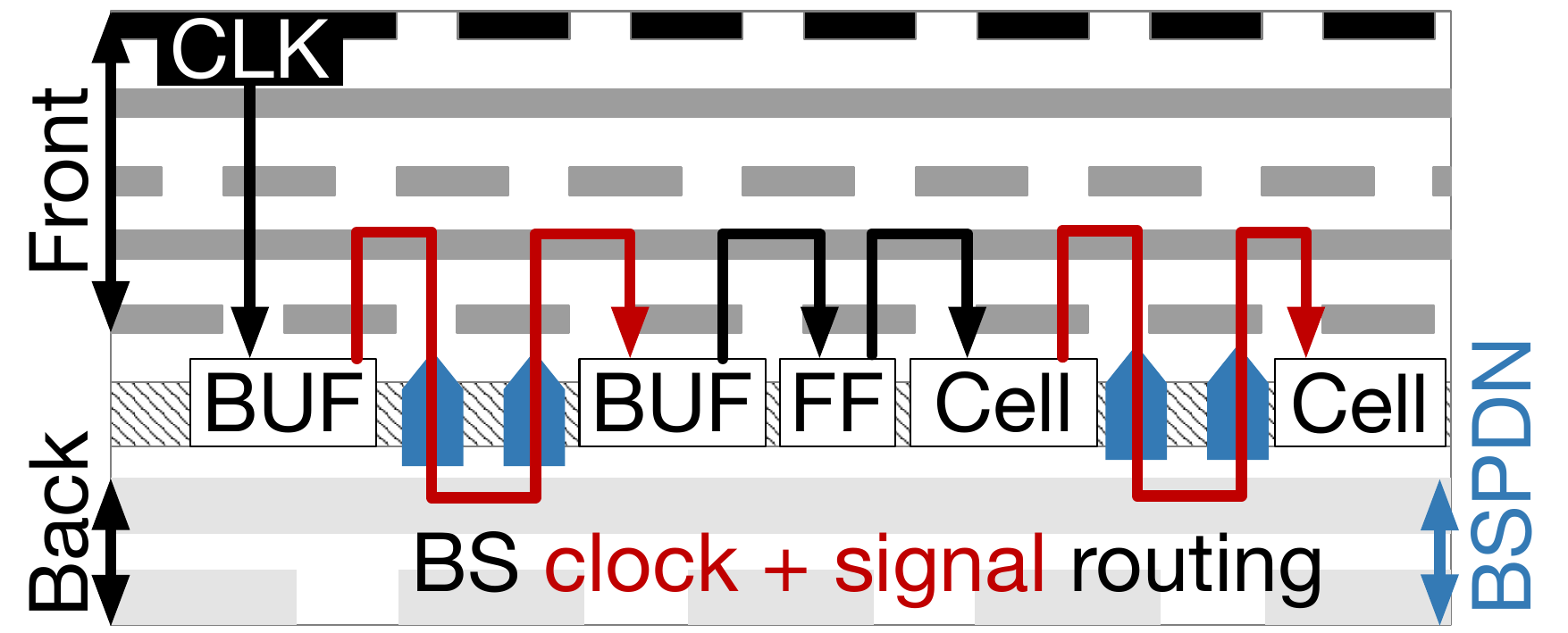}\label{fig:f1d}}
  \caption{Backside design for clock-signal co-optimization with limited backside resources.}
  \label{fig:main}  
\end{figure}

\begin{figure*}[t!]
	\centering
	\subfloat[BSPDN Structure]{
		\includegraphics[height=95pt]{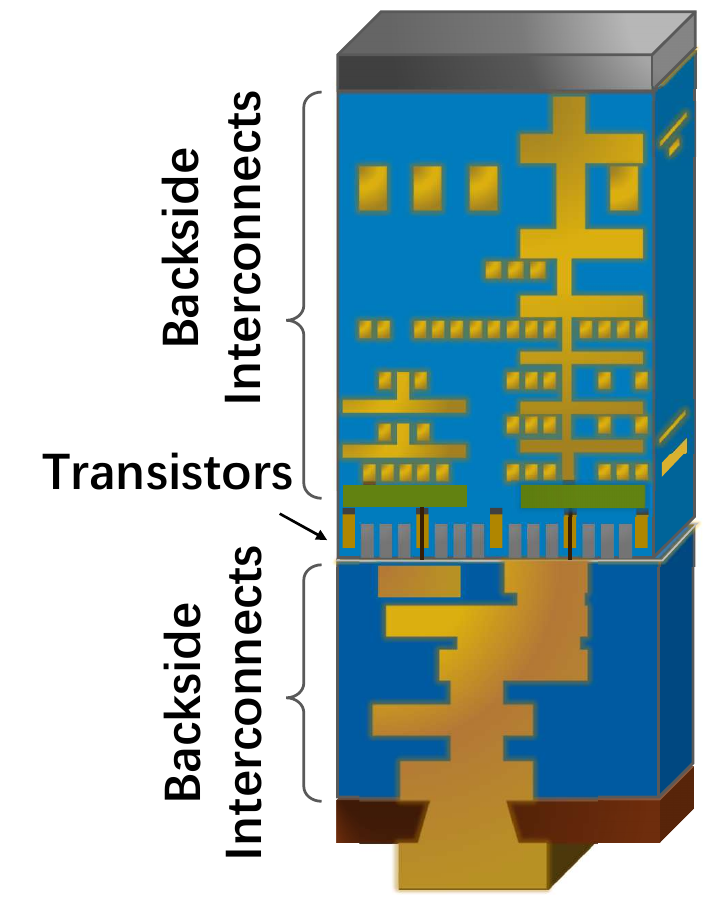}
		\label{fig:related-a}
	}
    \subfloat[Shielding with FSPDN]{
		\includegraphics[height=95pt]{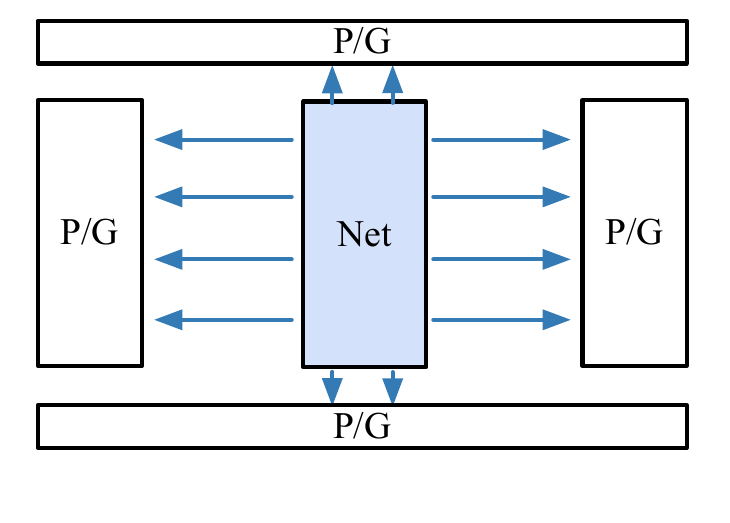}
		\label{fig:related-b}
	}
	\subfloat[BSPDN SI Problem]{
		\includegraphics[height=95pt]{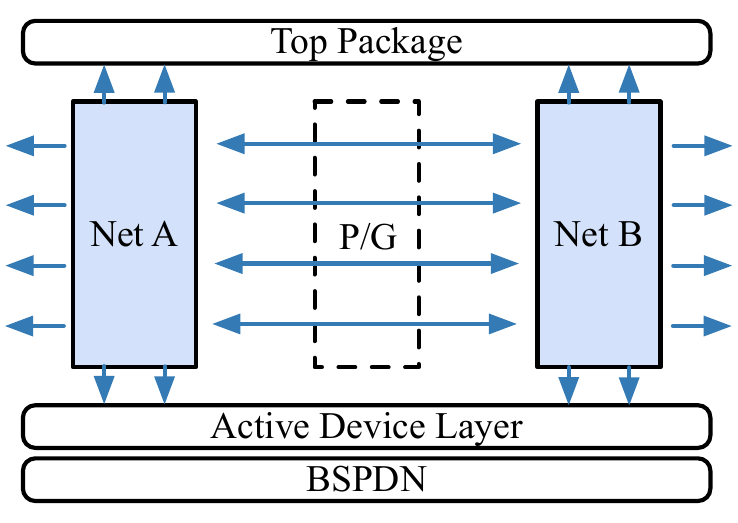}
		\label{fig:related-c}
	}  
    \subfloat[Crosstalk induced delay]{
		\includegraphics[height=95pt]{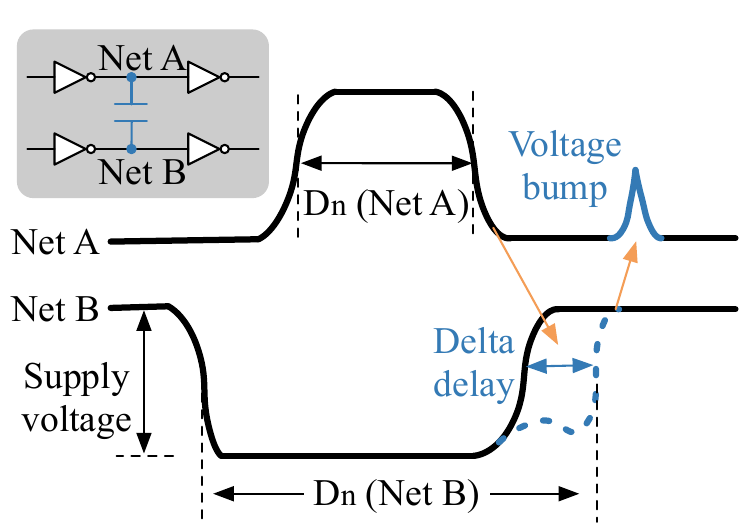}
		\label{fig:related-d}
	} 
    \caption{SI problem on BSPDN design.}
	\label{fig:related}
\end{figure*}

However, this architectural shift breaks the conventional separation between signal, clock, and power networks, demanding new cross-layer co-design methodologies~\cite{pdn,PDNLast,gnn,omni}. Current research has explored backside routing in progressively broader yet still isolated ways, as illustrated in~\Cref{fig:main}. Early BSPDN implementations focus exclusively on power delivery without exploiting residual backside routing resources~\cite{PDNLast}. Subsequent clock-specific approaches (\Cref{fig:f1b}) relocate clock networks using specialized backside buffer cells, demonstrating improvements in clock skew and latency~\cite{VLSI24,gnn,Lin25}. Other methods~\cite{bounding,bbox-1,eco,LimVLSI25} have explored the relocation of signal nets according to their geometry features (\Cref{fig:f1c}), with an assumption that the clock net is pre-constructed.

Collectively, these BSPDN approaches suffer from three critical limitations that fundamentally constrain achievable PPA gains. First, clock and signal optimization remain decoupled, so methods targeting only clocks neglect timing-critical signal paths, while geometric heuristics for signals often miss compact yet performance-critical nets due to weak correlation between timing criticality and geometric properties. Second, existing approaches usually presuppose fixed power grid configurations (\textit{e.g.,} predetermined metal stripe widths and pitches) prior to net selection. However, the backside net routing and PDN configuration are inherently dependent. PDN synthesis proceeds without knowledge of routing demands, while routing engines ignore PDN-induced blockages, fragmenting the design space and preventing discovery of design-specific optimal solutions that balance power integrity, routing congestion, and timing closure. Third, existing works overlook a newly emergent physical effect: removing frontside power grids eliminates electromagnetic shielding between signal nets, creating a "gravity-less" coupling environment where crosstalk increases 2--4$\times$~\cite{SI}. Without the confinement of frontside power straps, signal metals experience amplified noise propagation that can offset or even negate BSPDN's timing benefits~\cite{XT,C20,GraphCAD}. This insight motivates a co-optimization framework that exploits backside power layers as shields for SI-critical paths while jointly allocating backside resources across signal and clock networks.

To address these issues, we propose {CSCO}, a BSPDN-aware co-optimization framework that jointly determines the BSPDN topology and the allocation of signal and clock nets across frontside and backside layers (\Cref{fig:f1d}). Unlike prior approaches that fix PDN configurations before net selection, CSCO recognizes that optimal power grid structures depend on design-specific routing demands and explores both dimensions simultaneously. It employs fast surrogate modeling to navigate the coupled design space without costly full-routing iterations and exploits BSPDN shielding to protect SI-critical nets. This approach enables discovery of the better solutions that balance power integrity, signal integrity, and PPA metrics. Our key contributions are summarized as follows:
\begin{itemize}[leftmargin=4mm]
    \item {Unified Co-Optimization Framework:} The first methodology that concurrently synthesizes dual-side signal and clock networks under BSPDN constraints, achieving cross-layer integration of routing and power delivery planning.
    \item {Data-Driven Surrogate Modeling:} A scalable optimization scheme that employs low-rank statistical modeling to capture timing--resource interactions without costly full-routing evaluations.
    \item {Signal Integrity Mitigation via BSPDN Shielding:} Exploiting BSPDN’s electromagnetic properties to improve SI robustness.
    \item {Demonstrated Effectiveness:} On realistic CPU benchmarks, CSCO achieves significant improvement in PPA metrics while satisfying power and signal integrity constraints.
\end{itemize}

\section{Preliminaries}
\label{sec:pre}
Before presenting our method, we briefly introduce necessary background.

\subsection{Backside PDN}
BSPDNs use thick, low-resistivity backside metals connected to frontside buried power rails by sub-100-nm nTSVs~\cite{zhou2024vlsi,linqiu,ivr,lu2024first,A14-ppa}. This separation reduces power-distribution resistance by up to 30\% and frees frontside layers for signal routing~\cite{probe3,omni,dtco,stco}, as shown in~\Cref{fig:related-a}.

\minisection{Clock-Centric Backside Routing}
Existing approaches~\cite{VLSI24,Lin25,PDNLast,VLSI25} relocate entire clock networks using backside buffer cells and nTSVs, improving skew and latency through GNN-based endpoint classification~\cite{gnn} or analytical optimization~\cite{Lin25}. However, they ignore timing-critical signal routes and may migrate complete trees when only critical branches need optimization, wasting backside resources and nTSVs.

\minisection{Signal-Centric Backside Routing}
Signal-centric methods~\cite{bounding,bbox-1,eco,LimVLSI25} select nets using bounding-box size, pin count, or routing congestion. These geometric proxies can miss compact critical nets and favor large noncritical nets, producing suboptimal allocations. Net-level migration can also conflict with backside PDN structures, causing DRC violations that require post-routing fixes~\cite{PDNLast}.

\minisection{Lack of Clock-Signal Co-optimization}
Existing works optimize clock and signal routing sequentially~\cite{Lin25,gnn,bbox-1,eco,VLSI24}, despite their interdependence and competition for limited backside resources. Allocating more resources to clocks necessarily leaves fewer for signals, and vice versa. Clock tree synthesis determines signal-path timing budgets, whereas subsequent signal migration cannot influence clock routing. This separation prevents exploration of clock-signal resource-allocation tradeoffs.

\minisection{SI Degradation Induced by BSPDN}
Relocating the PDN to the backside removes frontside P/G grids that normally shield signal nets~\cite{SI}. As shown in~\Cref{fig:related-b,fig:related-c}, FSPDN strips maintain favorable self-to-coupling capacitance ratios, whereas BSPDN eliminates this shielding and increases direct coupling between adjacent nets. The resulting aggressor-victim interference causes delta delay $\Delta D$ and voltage bumps on timing-critical nets (\Cref{fig:related-d}). Backside signal routing can therefore exacerbate SI issues, requiring joint optimization of timing, routability, and signal integrity under limited backside resources.

\begin{figure}[t!]
    \centering
    \includegraphics[width=\linewidth]{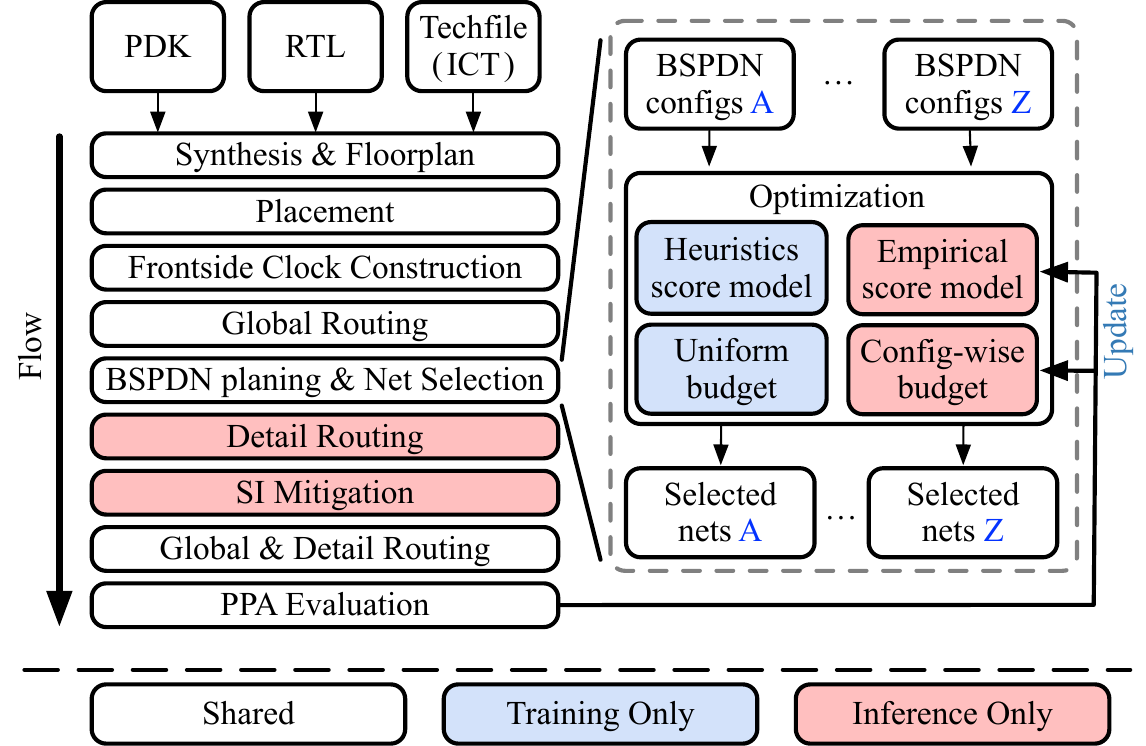}
    \caption{Overall design flow of CSCO.}
    \label{fig:pipeline}
\end{figure}

\subsection{Efficient Surrogate Model}

Machine-learning surrogates replace costly physical-design evaluations in EDA. Neural models predict routability~\cite{xie2018routenet}, pre-routing timing~\cite{barboza2019machine}, and slack~\cite{guo2022timing}, but typically require thousands of full-routing and signoff labels.
Gaussian-process Bayesian optimization~\cite{shahriari2015taking} is more data-efficient and used in analog synthesis~\cite{lyu2018batch}, but cubic scaling and continuous, low-dimensional assumptions limit high-dimensional binary allocation.
Factorization machines (FMs)~\cite{rendle2010factorization}, originally developed for recommendation systems, capture pairwise interactions through low-rank embeddings with $\mathcal{O}(nl)$ complexity, require little data, and naturally handle binary inputs. This compact parameterization suits costly signoff labels. Their success in materials and drug-discovery optimization~\cite{tamura2025blackboxoptimizationusingfactorization}, together with limited use in EDA, motivates our adoption.

\subsection{Evaluation Metrics}
\label{subsec:metrics}

We evaluate the practical impact of backside allocation on overall design performance.

\minisection{WNS and TNS}
Worst Negative Slack ($T_{\text{wns}}$) and Total Negative Slack ($T_{\text{tns}}$) are standard timing indicators; larger values imply better performance, enabling higher effective frequency.

\minisection{Delta-Delay and Bump Ratios}
Signal integrity is characterized using the delta-delay ratio and bump ratio:
\begin{equation}
r_{delay} = \frac{\Delta D}{D_d + D_n + \Delta D}, \quad
r_{bump} = \frac{V_{bump}}{V_{DD}}.
\end{equation}
Here, $D_d$ and $D_n$ are the intrinsic driver and net delays, and $V_{DD}$ is the supply voltage. Higher ratios indicate stronger SI-induced delay or voltage disturbance, and we additionally count how many nets exceed predefined thresholds for these two metrics.

\section{Methods}

We now introduce the proposed BSPDN-aware design framework, CSCO, which operates within the backside-aware design flow detailed in \Cref{subsec:flow}. As illustrated in \Cref{fig:pipeline}, CSCO augments the conventional 3D-IC implementation process by introducing data-driven optimization stages that jointly consider net partitioning, BSPDN planning, and timing robustness.

\minisection{Framework Overview}
CSCO operates in two major phases, training and {inference}, to transfer knowledge from empirical exploration to efficient design-space prediction. 
The training phase focuses on data collection and surrogate model learning, while the inference phase exploits the trained model for BSPDN-aware optimization and timing estimation. 

Within the framework, a \textit{key novelty} of CSCO lies in its BSPDN-aware formulation, which links PDN planning, net partitioning, and timing through a unified surrogate-driven optimization flow.

\subsection{Problem Formulation}
The goal of this stage is to determine the optimal allocation of all nets $\mathcal{N}$ in a given design under a specific BSPDN configuration. 
Let $\mathcal{N} = \{ s_1, \ldots, s_m, c_1, \ldots, c_n \}$, where $s_i$ and $c_j$ denote signal and clock nets, respectively. With above symbols, we formulate backside resource allocation as an optimization problem under BSPDN-dependent constraints:
\begin{equation}
    \text{Eval}(\vx) = T_{\text{wns}}(\vx) + \lambda\, T_{\text{tns}}(\vx),
    \label{eq:eval}
\end{equation}
where $T_{\text{wns}}$ and $T_{\text{tns}}$ denote the worst and total negative slack, respectively, and $\lambda$ controls their relative importance.
The design variable is a binary vector 
\begin{equation}
    x_i \in \{0,1\}, \quad \forall i \in [1, m+n],
    \label{eq:binary}
\end{equation}
where $x_i{=}1$ indicates that net $i$ is assigned to the backside, otherwise the net is assigned to the frontside.

\begin{figure}[t]
    \centering
    \includegraphics[width=\linewidth]{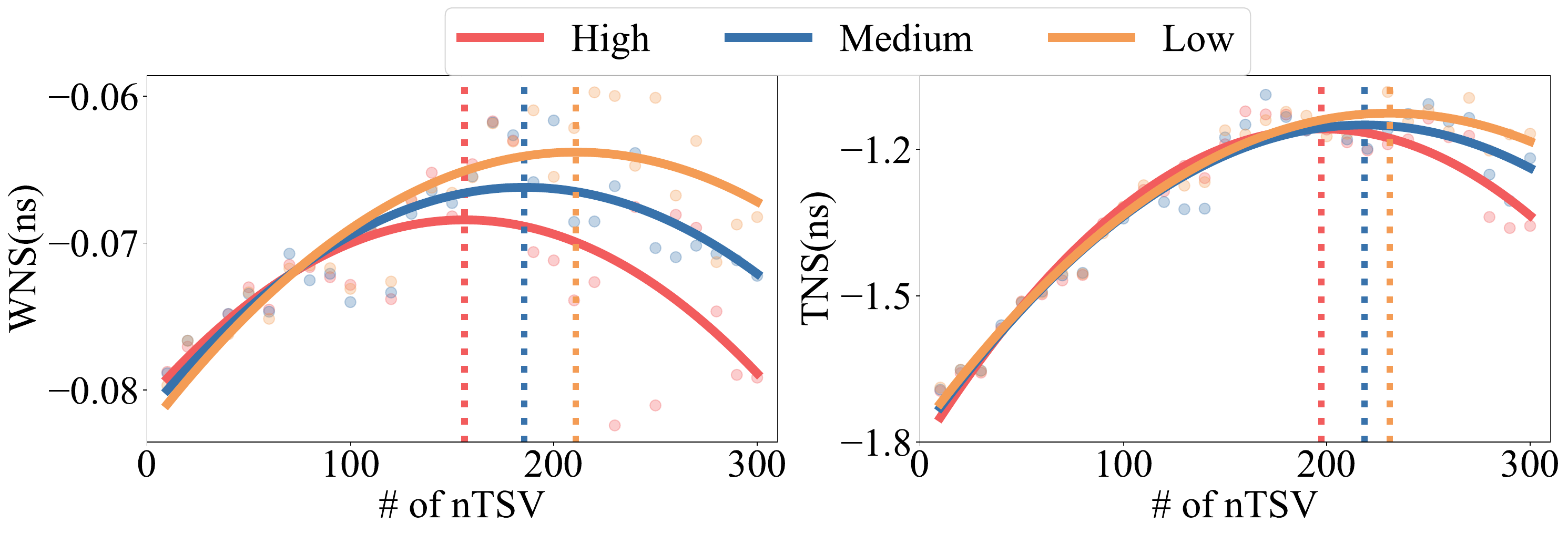}
    \caption{Timing performance (WNS and TNS) as a function of nTSVs count under different BSPDN densities. 
    }
    \label{fig:budget}
\end{figure}

\begin{figure}[t]
    \centering

    \includegraphics[width=\linewidth]{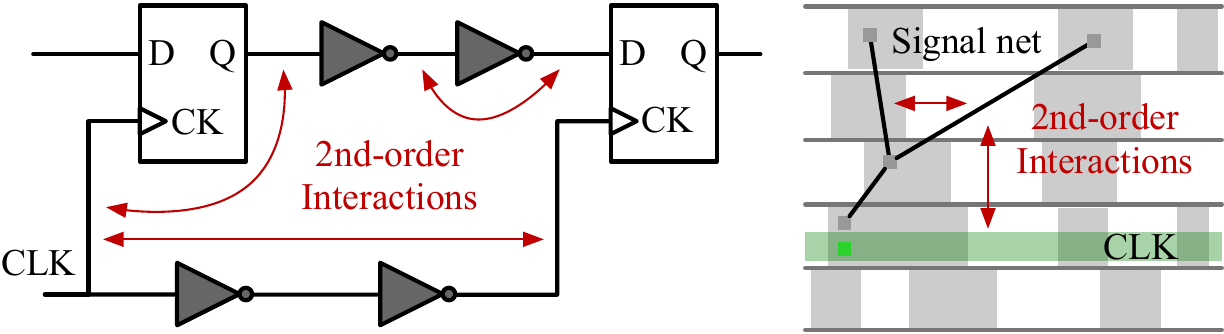}
    \caption{The second-order relationship between nets.}
    \label{fig:2order}
\end{figure}

This optimization problem is challenging for the following three reasons. 
First, evaluating any candidate $\vx$ requires full routing and timing analysis, making each trial extremely expensive. 
Second, the search space is a high-dimensional 0–1 domain with hundreds of thousands of variables, rendering brute-force or classical combinatorial methods infeasible. 
Third, extending this basic formulation to be explicitly BSPDN-aware is nontrivial: the BSPDN configuration simultaneously impacts IR-drop, routing resources, and timing, and must be incorporated through a compact yet physically meaningful constraint, which we introduce next via a configuration-dependent nTSV budget model.

\minisection{BSPDN-Dependent Constraints and Budget Determination}
Unlike prior works that treat nTSV resources as fixed, we explicitly model BSPDN dependence, allowing power-network density to directly affect routing capacity and timing. 
A denser BSPDN enhances IR-drop robustness but limits routing resources, so we capture this trade-off through a global nTSV budget constraint:
\begin{equation}
    \sum_{i=1}^{m+n} t_i\, x_i \le B,
    \label{eq:budget}
\end{equation}
where $t_i$ is the number of nTSVs required by net $i$ (equal to its pin count). 
Timing improves as $B$ increases until a configuration-dependent threshold, beyond which excessive nTSVs degrade routability, as shown in \Cref{fig:budget}.  

The budget $B$ is determined empirically by sweeping nTSV allocations and recording corresponding PPA outcomes. 
The inflection point of the performance–nTSV curve is selected as the final budget, which remains fixed during inference for consistent evaluation across configurations.  
This BSPDN-aware formulation links power-delivery planning with timing optimization, bridging two stages traditionally treated independently in 3D-IC design flows.

However, a direct optimization on \Cref{eq:eval,eq:binary,eq:budget} is computationally prohibitive due to expensive routing and timing analysis. 
To address this challenge, we employ a data-efficient surrogate model to approximate $\text{Eval}(\vx)$ 
and enable scalable optimization.


\subsection{Surrogate Model}
\label{subsec:surrogate}

The true evaluation function $\text{Eval}(\vx)$ requires full routing and timing analysis, 
making data collection extremely expensive. 
Such scarcity renders deep-learning surrogates impractical, 
motivating the use of a lightweight and interpretable model that can learn from limited samples.

\minisection{Low-Rank Interaction Modeling} To efficiently approximate the objective function \Cref{eq:eval} with limited data, we employ a second-order FM model~\cite{rendle2010factorization}. Given a binary partition vector $\vx \in \{0,1\}^{m+n}$ indicating backside assignment, the $k$-order FM prediction is defined as: \begin{equation} \hat{y}(\vx) = w_0 + \vw^\top \vx + \sum_{d=2}^{k} \sum_{i_1 < \cdots < i_d} \langle \vv_{i_1}, \ldots, \vv_{i_d} \rangle \prod_{j=1}^{d} x_{i_j}, \label{eq:high_order_fm} \end{equation} where $w_0 \in \mathbb{R}$ is a global bias, $\vw \in \mathbb{R}^{m+n}$ models first-order (individual-net) effects, $\vv_i \in \mathbb{R}^{l}$ is a latent embedding vector representing the hidden timing characteristics of net $i$ and the operator $\langle \vv_{i_1}, \ldots, \vv_{i_d} \rangle = \sum_{f=1}^{l} \prod_{j=1}^{d} v_{i_j,f}$ captures $d$-way interactions. All $\vv_i$ are assembled into the latent matrix $\mV = [\vv_1, \vv_2, \ldots, \vv_{m+n}]^\top$, so that $\{w_0, \vw, \mV\}$ constitute all learnable parameters of the model.

In this work, we adopt the second-order case ($k{=}2$) for several circuit-level considerations: 1) most timing dependencies arise from \textit{pairwise} interactions, including between signal nets connected to the same cell, and between clock and signal nets along critical paths, as illustrated in \Cref{fig:2order}. Higher-order correlations contribute marginally yet increase data complexity. Thus, a second-order FM effectively captures the dominant physical dependencies while remaining data efficient. 2) Compared with neural surrogates, FM offers greater interpretability and stability under limited training data, making it particularly suitable for BSPDN-aware optimization. 3) Furthermore, once trained, the FM model generalizes across different BSPDN configurations without retraining, since it encodes intrinsic inter-net relationships rather than configuration-specific parameters. This cross-configuration consistency allows the same surrogate to guide optimization under varying PDN densities.

Additionally, a standard simplification~\cite{rendle2010factorization} for $k{=}2$ reduces computational complexity from $\mathcal{O}((m+n)^2 l)$ in \Cref{eq:high_order_fm} to $\mathcal{O}((m+n)l)$:
\begin{equation}
    \hat{y}(\vx) = w_0 + \vw^\top \vx 
    + \tfrac{1}{2}\!\left(\|\mV^\top\vx\|_2^2 - \|\text{Diag}(\vx)\mV \|_F^2\right),
    \label{eq:vectorized_fm}
\end{equation}
which significantly improves efficiency and is well-suited for GPU parallelization.

\minisection{Cold-Start Data Construction}
Before training the surrogate model, we require an initial dataset for cold-start learning. 
We propose an empirical {selection criterion} to identify nets most likely to impact timing when migrated to the backside. 
For each net type (signal or clock), we extract five post-global-routing metrics that are lightweight to compute and available without detailed routing, as summarized in \Cref{tab:criterion}.

After normalizing each metric to $[0,1]$, we compute a composite score quantifying the timing improvement potential of each net:
\begin{align}
\text{Score}_{\text{sig}} = &
    \alpha_1 \tilde{t}_{\text{tran}} +
    \alpha_2 \tilde{n}_{\text{load}} +
    \alpha_3 \tilde{C}_{\text{est}} +
    \alpha_4 \tilde{N}_{\text{path}}
    -\, \alpha_5 \tilde{r}_{\text{first}},  \label{eq:score-1}\\
\text{Score}_{\text{clk}} =&
    \beta_1 \tilde{t}_{\text{tran}} +
    \beta_2 \tilde{n}_{\text{load}} +
    \beta_3 \tilde{N}_{\text{crit}} -
    \beta_4 \tilde{t}_{\text{arr}}
    -\, \beta_5 \tilde{r}_{\text{first}}, 
\label{eq:score-2}
\end{align}
where $(\tilde{\cdot})$ denote normalized values and $\alpha_i$, $\beta_i$ are weighting coefficients (set equally in this work). 
A higher score indicates a stronger likelihood that the corresponding net will contribute positively to timing performance if allocated to the backside.
\begin{table}[t!]
    \centering
    \caption{Normalized post-global-routing metrics used for cold-start net selection.}
    \setlength\tabcolsep{3.2pt}
    \fontsize{7}{9}\selectfont 
    \begin{tabular}{cccc}
    \toprule
    Symbol (Unit) & Definition  & Range & Net Type \\
    \midrule
    $\tilde{t}_{\text{tran}}$ (ps) & Transition delay  & $\ge 0$ & Both \\
    $\tilde{r}_{\text{first}}$ (N/A) & Rank of first critical path &  $\{1,\dots,m{+}n\}$ & Both \\
    $\tilde{n}_{\text{load}}$ (N/A)& Number of loads  & $\mathbb{Z}^+$ & Both \\
    \midrule
    $\tilde{C}_{\text{est}}$ (fF)& Estimated net capacitance & $\ge 0$ & Signal \\
    $\tilde{N}_{\text{path}}$ (N/A) & Occurrence count in all paths &$\{1,\dots,m{+}n\}$ & Signal \\
    \midrule
    $\tilde{t}_{\text{arr}}$ (ns) & Arrival time & $\ge 0$ & Clock \\
    $\tilde{N}_{\text{crit}}$ (N/A)& Occurrences in top-200 critical paths  & $\{0,\dots,200\}$ & Clock \\
    \bottomrule
    \end{tabular}
    
    \label{tab:criterion}
\end{table}

This criterion efficiently constructs the training dataset by focusing computation on timing-critical nets, 
thereby reducing routing iterations required for model initialization.


\subsection{Optimization and Inference}
\label{subsec:opt}
With the surrogate model established, optimization proceeds in two phases: 
the {training phase} for generating diverse samples, 
and the {inference phase} for obtaining the BSPDN-aware partitioning.

\minisection{Training Phase: Greedy + Exploration}
During training, we aim to collect high-quality yet diverse partitioning samples. 
Nets are first ranked by their normalized utility, defined as the empirical score 
(\Cref{eq:score-1,eq:score-2}) divided by the required number of nTSVs $t_i$.  
We then perform greedy selection under a preset clock/signal ratio until the total nTSV budget is reached. 
To enhance diversity, a fixed fraction of selected nets is randomly replaced in each run, enabling broader exploration and improving surrogate coverage.

\minisection{Inference Phase: FM-Guided Greedy + Simulated Annealing}
In inference, the trained FM model provides a fast estimation of timing performance. 
For each net, we compute its importance by combining the first-order term 
and its pairwise interactions in the FM output, then divide by $t_i$ to reflect 
its benefit per backside cost:
\begin{equation}
    I_i = \frac{w_i + \sum_{j \ne i} \langle \vv_i, \vv_j \rangle x_j}{t_i},
    \label{eq:importance}
\end{equation}
where $I_i$ denotes the importance of net $i$, $w_i$ is its first-order coefficient, 
$\langle \vv_i, \vv_j \rangle$ represents the pairwise interaction from the FM model, 
and $t_i$ is the number of nTSVs required by the net.
We again perform greedy selection under the clock/signal ratio until 
the total budget is filled, yielding an initial feasible solution. To further refine this solution, we adopt {simulated annealing} (SA) directly on the surrogate.  
At each iteration, a few nets are swapped between dies to obtain a new candidate $\vx'$, 
and the update is accepted with probability:
\begin{equation}
    P_{\text{accept}} = 
    \min\!\left(1, 
    \exp\!\left(\frac{\hat{y}(\vx') - \hat{y}(\vx)}{T}\right)\right),
    \label{eq:sa_accept}
\end{equation}
where $\hat{y}(\cdot)$ is the FM-predicted performance and $T$ is the temperature 
that gradually cools down (e.g., $T_{t+1} = \eta T_t$, $\eta \in (0,1)$).  
The optimization stops when no further improvement is observed or a preset iteration limit is reached.  
The final partitioning $\vx^*$ is used for full routing and PPA evaluation.
\begin{figure}[t]
  \centering
  \subfloat[Delta delay ratio]{\includegraphics[width=0.49\linewidth]{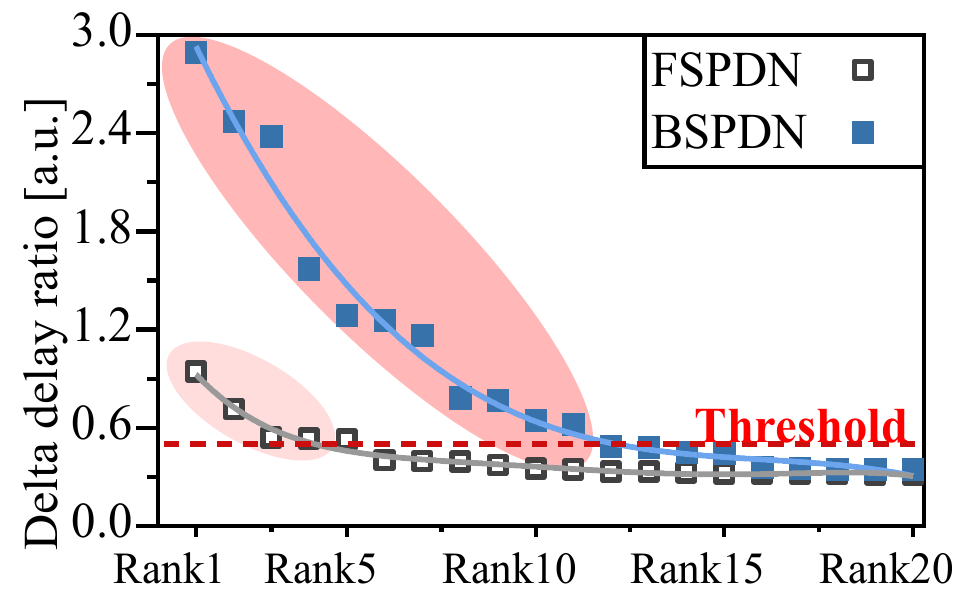}\label{fig:sifix-a}}
  \hfill
  \subfloat[Voltage bump]{\includegraphics[width=0.49\linewidth]{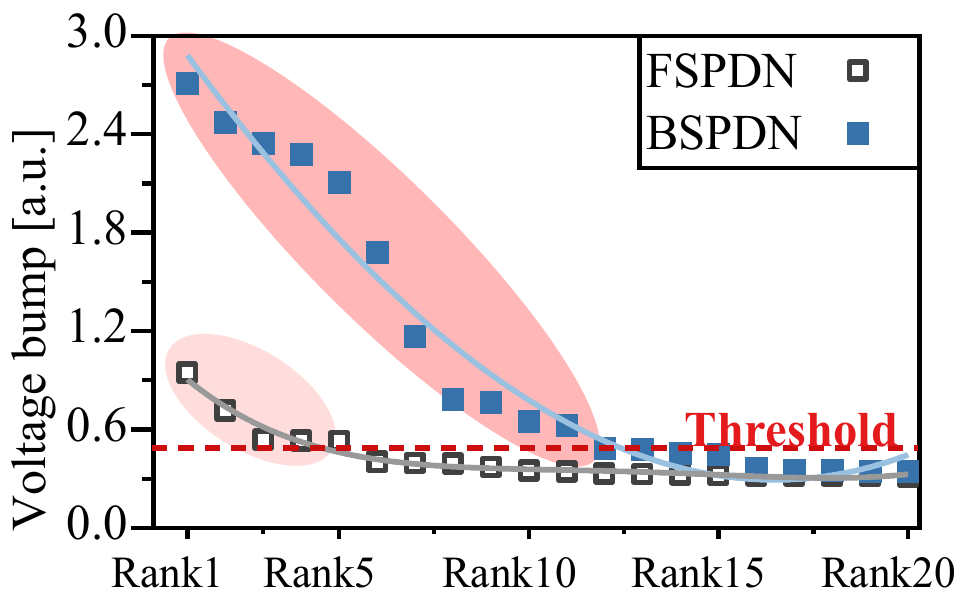}\label{fig:sifix-b}}
  \caption{Crosstalk impacts on nets: voltage bump and delta-delay at 4\si{\giga\hertz} operating frequency.}
  \label{fig:sifix}
\end{figure}

\begin{table*}[t!]
    \centering
    \caption{Performance metrics comparison across all benchmarks at 2\si{\giga\hertz} target frequency. \textbf{Bold} values indicate the best results. FS denotes frontside PDN with frontside routing. BSPDN+FR denotes BSPDN with frontside routing.}    
    \label{tab:ppa}         
    \fontsize{7}{9}\selectfont 
    \setlength\tabcolsep{1.2pt}
    \begin{tabular}{c|ccccc|ccccc|ccccc}
        \toprule
        & \multicolumn{5}{c|}{{\textbf{SHA256 }(\#Cell = 7.9k)}} & \multicolumn{5}{c|}{{\textbf{JPEG} (\#Cell = 182k)}} & \multicolumn{5}{c}{{\textbf{ARM Cortex-A7} (\#Cell = 108k)}} \\ 
        &\textbf{FS}  & \textbf{BSPDN+FR} & \textbf{DAC24}\cite{gnn} &  \textbf{DAC25}\cite{Lin25} & {\textbf{CSCO}} 
        &\textbf{FS}  & \textbf{BSPDN+FR} & \textbf{DAC24}\cite{gnn} &  \textbf{DAC25}\cite{Lin25} & {\textbf{CSCO}}
        &\textbf{FS}  & \textbf{BSPDN+FR} & \textbf{DAC24}\cite{gnn} &  \textbf{DAC25}\cite{Lin25} & {\textbf{CSCO}} \\ \midrule        
        \textbf{{Utilization (\%)}} & 82.91 & 82.36&82.36&82.36&82.36& 78.24 & 75.50 & 75.50& 75.50& 75.50& 78.16 & 74.15& 74.15& 74.15& 74.15 \\            
        {\textbf{Area (\si{\micro\meter^2})}} & 1319 & 1311 & 1311& 1311& 1311& 31285 &32280&32280&32280&32280& 14267&13491 &13491&13491&13491\\ 
        {\textbf{\#Metal (BS+FS)}} & 0+6 & 3+6 & 3+6& 3+6& 3+6 & 0+7 & 3+7 & 3+7& 3+7& 3+7 & 0+8 & 3+8 & 3+8& 3+8& 3+8\\     
       
        {\textbf{\#nTSV}} & N/A & N/A & 270 & 194 & 230  & N/A & N/A & 396 & 768 & 800 & N/A & N/A & 316 & 634 & 600 \\ 
     
        {\textbf{Clock Skew (\si{\pico\second})}} & 24.17 & 27.94 & 27.02 & 19.72 & 32.09 & 112.16 & 35.37 & 31.93 & 24.86 & 34.17 & 217.67 & 95.06 & 66.79 & 52.97 & 43.49 \\ 
        {\textbf{Clock Latency (\si{\pico\second})}} & 83.29 & 75.31 & 64.63 & 51.47 & 75.13 & 226.26 & 149.31 & 112.11 & 79.31 & 124.46 & 456.98 & 206.51 & 147.67 & 123.87 & 135.80 \\  
        {\textbf{Wirelength (\si{\m})}} & 0.05 & 0.04 & 0.04 & 0.04 & 0.04 & 0.81 & 0.70 & 0.69 & 0.68 & 0.69 & 1.47 & 1.37 & 1.38 & 1.38 & 1.39 \\  
          {\textbf{Total Power (\si{\mW})}} & 7.41 & 7.17 & {6.77} & 7.09 & {7.05} & 152.92 & 143.42 & 149.01 & 137.44 & 140.21 & 140.97 & 131.93 & 122.67 & 120.01 & 142.68 \\ 
          {\textbf{IR-drop (\si{\mV})}} & 62.54 & 14.13 & 14.32 & 14.25 & 14.35 & 55.95 & 23.79 & 24.09 & 24.02 & 25.99 & 170.02 & 17.88 & 18.93 & 18.03 & 18.58 \\
          \hline
        {\textbf{T}$_{\text{\textbf{wns}}}$ (\si{\nano\second})} & -0.196 &-0.111 & {-0.094} & -0.106 & \textbf{-0.061}  & -0.150 & -0.126 & -0.122  & -0.125 & \textbf{-0.077} & -0.194  & -0.081 & -0.092 & -0.106 & \textbf{-0.011} \\ 
        {\textbf{T}$_{\text{\textbf{tns}}}$ (\si{\nano\second})} & -4.06 & -1.96 & -2.04 & -2.14 & \textbf{-1.09} & -28.53 & -25.25 & -24.44 & -25.06 & \textbf{-7.66} & -1.756 & -1.637 & -1.579 & -1.499 & \textbf{-1.139} \\ 
        {\textbf{Eff. Freq} (\si{\giga\hertz})} & 1.50 & 1.64 & 1.70 & 1.65 & \textbf{1.78} & 1.54 & 1.61 & 1.63 & 1.60 & \textbf{1.73} & 1.44 & 1.72 & 1.69 & 1.65 & \textbf{1.96} \\   
        \textbf{\#$r_{delta}>$ 40\%} & 3 & 8 & 8 & 8 & \textbf{2} & 3 &  16 & 16 & 16 & \textbf{4} & 7 & 17 & 17 & 17 & \textbf{2} \\ 
        \textbf{\#$r_{bump}>$ 30\%} & 1 & 3 & 3 & 3 & \textbf{0} & 2 &12  & 12 & 12 & \textbf{2} & 5 & 27 & 26 & 27 & \textbf{3} \\  
        \bottomrule
    \end{tabular}
\end{table*}

\begin{table}[t]
        \caption{Technology specifications and design rules.}
        \label{tab:parameters}
        \centering  
        \setlength\tabcolsep{1.3pt}
        \fontsize{7}{9}\selectfont 
        \begin{tabular}{l l l |l l l}
            \toprule     
            \textbf{BS} & \textbf{Width/Spacing(\si{\micro\meter})} & \textbf{Pitch(\si{\micro\meter})} & \textbf{FS} & \textbf{Width/Spacing(\si{\micro\meter})} & \textbf{Pitch(\si{\micro\meter})} \\
            \midrule
            \textbf{BSM0}     & 0.053 / 0.028 & 0.04  & \textbf{M0}              & 0.020 / 0.020 & 0.04  \\
            \textbf{BSM1}     & 0.125 / 0.375 & 0.5   & \textbf{M1}              & 0.037 / 0.020 & 0.06  \\
            \textbf{BSM2}     & 0.5 / 1.5     & 2     & \textbf{M2/3}            & 0.020 / 0.020 & 0.04 \\
            \textbf{BSM3}     & 0.5 / 1.5     & 2     & \textbf{M4-M8}    & 0.038 / 0.038 & 0.08  \\
            \bottomrule
        \end{tabular}

\end{table}

This two-phase design balances exploration and exploitation: 
the training phase diversifies data for surrogate learning, 
while the inference phase leverages the surrogate to efficiently refine 
the final BSPDN-aware allocation.

\subsection{Signal Integrity Mitigation}
Based on post-detailed-routing analysis of our initial BSPDN and FSPDN designs, we observe notable differences in signal integrity behavior. As shown in \Cref{fig:sifix}, BSPDN exhibits 2–4× higher voltage bumps and delay variations than FSPDN. The exponential decay trend in both metrics indicates that only a subset of nets is inherently more vulnerable to SI degradation. The red-shaded regions in \Cref{fig:sifix} mark the performance gap between the two PDN schemes, with the red dashed lines denoting the 0.5 ratio safety threshold. In our methodology, nets with voltage bump or delta-delay ratios exceeding 40\% are flagged as SI-critical. These nets are ranked by severity and explicitly reassigned to backside routing tracks, where stronger electromagnetic shielding can effectively suppress excessive interference, while preserving global routing balance and timing closure.

This SI optimization stage is designed to operate independently from the primary net selection in our flow, since SI degradation is driven more by routing congestion and net density than by timing metrics like WNS or TNS. Accordingly, we reserve backside routing resources for SI-critical nets, which proves more resource-efficient than extending BSPDN power rails into frontside metal layers.

\section{Experimental Results}
This section presents experiments on representative designs to validate the effectiveness of the proposed CSCO method. Both quantitative and qualitative results are reported.

\subsection{Experiment Settings}

\minisection{Technology Specifications}
Our BSPDN uses nTSVs to connect buried power rails (BPRs) to backside metal layers~\cite{linqiu}. BPRs reduce standard-cell height by eliminating conventional VDD/VSS BEOL layers above the FEOL. \Cref{tab:parameters} summarizes frontside and backside metal-stack parameters.

\minisection{Backside-Aware Design Flow}
\label{subsec:flow}
Since commercial implementation and signoff tools do not natively support double-sided flows with explicit backside metal layers, we developed a custom Interconnect Technology file (ICT) defining a backside-optimized BSM1--BSM3 BEOL stack and generated corresponding QRC files for parasitic extraction. TCAD simulations were translated into SPICE-compatible nTSV models with $C_{\text{nTSV}} = 0.444\,\text{fF}$ and $R_{\text{nTSV}} = 23\,\Omega$.

The flow iterates frontside and backside implementation with PDN-aware backside routing, parasitic extraction and merge, and signoff timing. It provides parasitic and timing feedback while exploiting backside routing resources without major modifications to established flows.

\minisection{Benchmark Designs}
We evaluate SHA256 (172 clock nets and 7.8k signal nets), JPEG (3.5k clock nets and 180k signal nets)~\cite{bs}, and ARM Cortex-A7 (1.5k clock nets and 116k signal nets)~\cite{arm}, a power-efficient CPU for mobile and embedded systems. All designs are implemented in 7\,nm technology ($V_{\text{DD}}=0.75$\,V) using Cadence Genus~\cite{genus} and Innovus~\cite{Innovus}. Power integrity is evaluated with an in-house IR-drop tool~\cite{IEDM22}, and timing is signed off using Synopsys PrimeTime SI~\cite{PT}.

\minisection{Implementation}
CSCO is implemented in Python 3.8 and evaluated on a Rocky Linux 8.10 cluster with two Xeon Platinum 8480+ CPUs (112 cores) and two NVIDIA L40S GPUs.

\minisection{Hyper-parameters}
We set the loss weight $\lambda=0.1$ and latent embedding dimension $l=12$. The FM is trained using MSE and SGD (batch size 32, learning rate $10^{-4}$, and $5\times10^3$ update steps), with 20\% of greedily selected nets randomly replaced for exploration. We collect 200 samples for SHA256 and 500 for both JPEG and ARM. During inference, simulated annealing starts at $T=10^{-5}$, cools with $\eta=0.995$, and stops at $10^{-6}$. Clock and signal net budgets are fixed at 30\% and 70\%, respectively.

\begin{figure}[t]
	\centering
	\subfloat[Signal net on critical path]{\includegraphics[width=4.15cm,height=3cm]{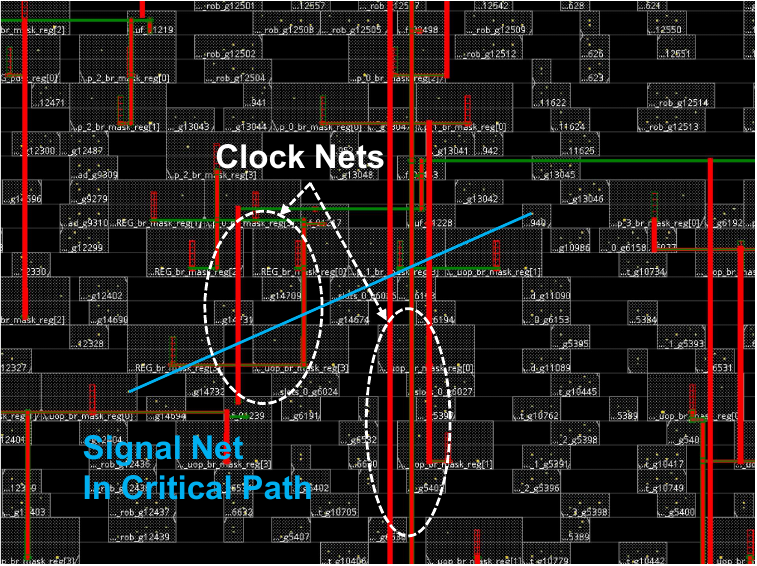}\label{fig:pdnvis-a}}\hfill
    \subfloat[Clock net on max-latency path]{\includegraphics[width=4.15cm,height=3cm]{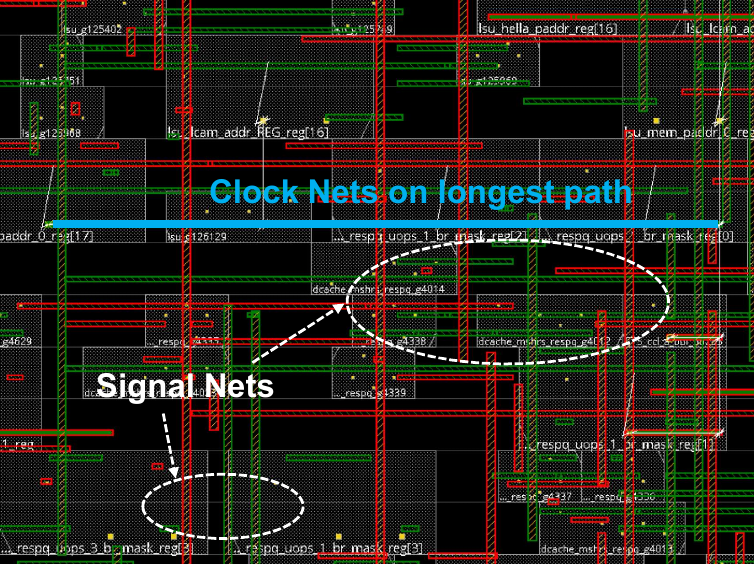}
	    \label{fig:pdnvis-b}}
	    \\
    \subfloat[Results on BSPDN-1]{\includegraphics[height=4cm]{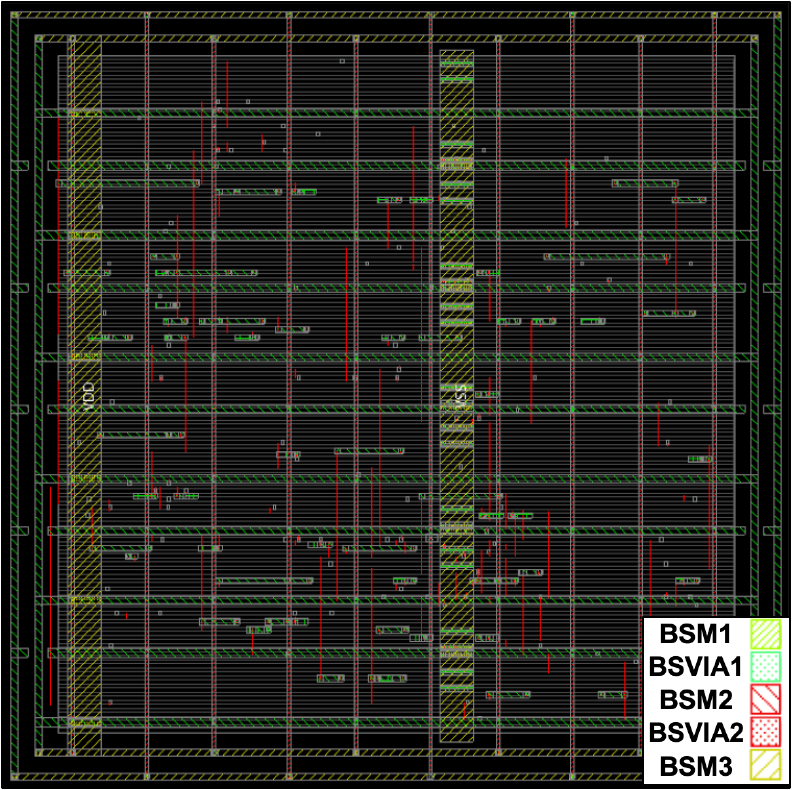}\label{fig:pdnvis-c}}\hfill
    \subfloat[Results on BSPDN-3]{\includegraphics[height=4cm]{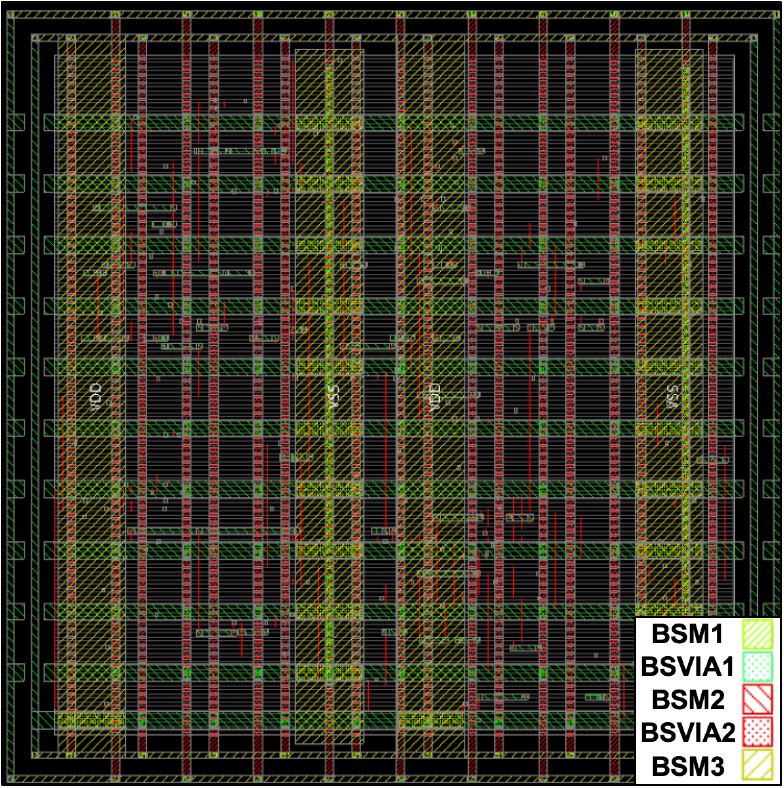}\label{fig:pdnvis-d}}
    \caption{Routing critical nets through backside and CSCO results under various BSPDN configurations.}
    \label{fig:pdnvis}    
\end{figure}

\subsection{PPA Evaluations}

In addition to the metrics from \Cref{subsec:metrics}, we report nTSV count, clock skew, wire length, total power, effective frequency, and IR-drop. \Cref{tab:ppa} compares CSCO across three benchmarks with baselines and two competitors, DAC24~\cite{gnn} and DAC25~\cite{Lin25}. FSPDN+FR and BSPDN+FR perform frontside-only routing with FSPDN and BSPDN, respectively; DAC24 applies GNN-based backside net assignment, whereas DAC25 routes critical-path clock nets on the backside.

CSCO achieves the best timing across all methods, improving WNS by up to 85\% on ARM Cortex-A7 and by over 60\% on average, and TNS by 35--70\%. These gains increase the effective frequency to 1.78\si{\giga\hertz} for SHA256 (20\% over FSPDN) and 1.96\si{\giga\hertz} for ARM (36.1\% over the baseline). CSCO also uses fewer nTSVs than DAC25 with negligible area overhead (e.g., 74.15\% utilization for ARM), demonstrating an effective balance between performance and resource efficiency.

\minisection{Visualization}
\Cref{fig:pdnvis-a,fig:pdnvis-b} visualizes critical-path signal and clock nets. Frontside congestion limits their routing efficiency, whereas backside routing expands the solution space for timing optimization.

\begin{figure}[t]
  \centering  
  \subfloat[Training (blue) and inference (red) solutions.]{\includegraphics[height=2.9cm]{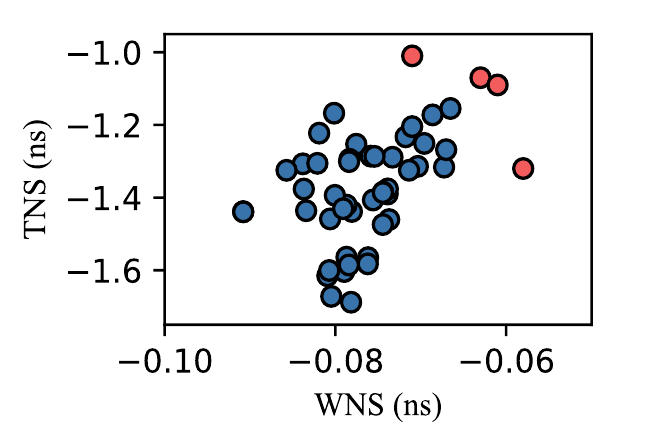}\label{fig:pareto}}
  \hfill
  \subfloat[Delta delay before and after SI mitigation.]{\includegraphics[height=2.9cm]{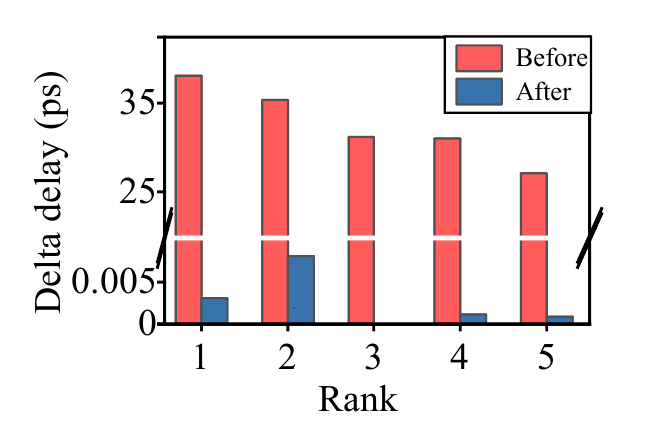}\label{fig:delaybar}}
  \caption{Solution quality and SI analysis.}
\end{figure}

\Cref{fig:pdnvis-c,fig:pdnvis-d} shows CSCO's backside routing results on SHA256 under two BSPDN configurations, illustrating how available backside resources affect routing patterns and utilization. \Cref{fig:pareto} compares heuristic and empirical-model solutions, showing that CSCO extends the Pareto frontier toward better solutions.

\subsection{Surrogate Model}

To understand how CSCO works, we evaluate the proposed surrogate FM model across multiple dimensions here.

\begin{figure}[t]
  \centering
  \subfloat[Objective prediction.]{\includegraphics[height=2.9cm]{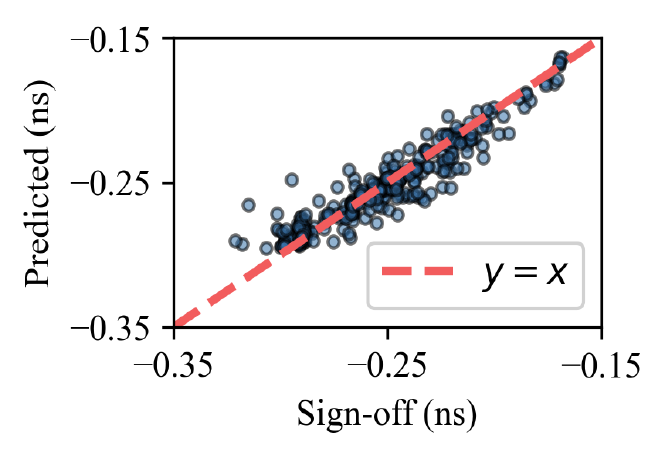}\label{fig:surrogate-a}}
  \hfill
  \subfloat[Effect of training size.]{\includegraphics[height=2.9cm]{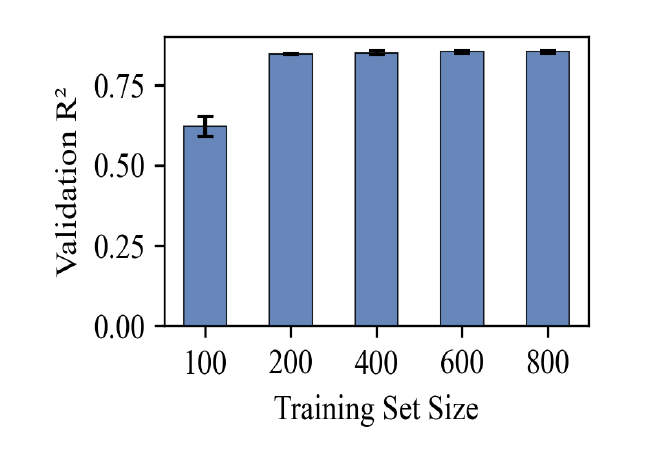}\label{fig:surrogate-b}}
  \caption{Surrogate model evaluation.}
  \label{fig:surrogate}
\end{figure}

\minisection{Accuracy and Consistency}
The surrogate FM in \Cref{subsec:surrogate} serves as a fast proxy for sign-off timing. On 337 held-out backside routing allocations from SHA256, its predictions correlate closely with the sign-off objective, achieving an $R^2$ of 0.858 and an RMSE of $0.0141$\,ns; the latter is small relative to the approximately $0.17$\,ns baseline objective. The correlation is particularly strong in the upper-right region of \Cref{fig:surrogate-a}, where better-timing solutions lie, making the model suitable for search guidance. It also preserves candidate rankings, with top-$K$ overlaps of 100\%, 90\%, and 85\% at $K=5$, 10, and 20, respectively.
\minisection{Training Robustness}
We evaluate validation $R^2$ on SHA256 with different training-set sizes, reporting the mean and standard deviation over three runs in \Cref{fig:surrogate-b}. At 200 samples, the consistently small variation indicates robustness to data sampling and optimization randomness. The model converges rapidly and generalizes well with only a few hundred samples, reducing calls to the expensive sign-off engine and accelerating co-optimization.

\minisection{Efficiency}
Training completes within 15 minutes on a single GPU, and each backside routing assignment is evaluated in under 30~\si{\milli\second}, compared with approximately 323.4 seconds per iteration for PrimeTime sign-off. Replacing sign-off analysis with the surrogate thus provides over $10{,}000\times$ faster objective evaluation. Moreover, one model applies to all BSPDN configurations of a given design without retraining, further reducing computational cost.

\begin{figure}[t]
  \centering
  \subfloat[Frontside routing]{\includegraphics[height=2.45cm]{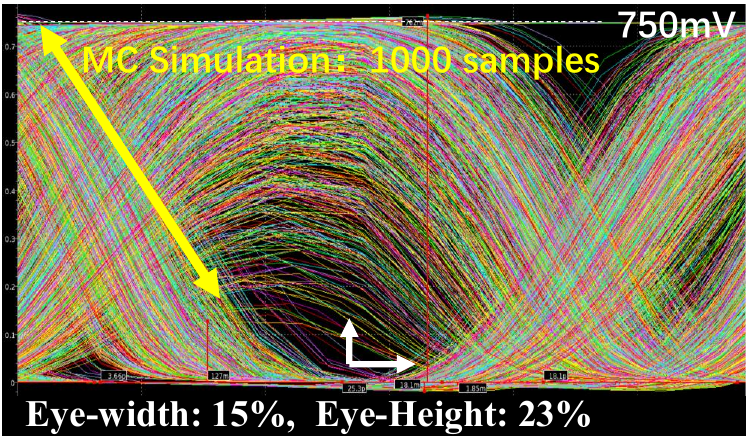}\label{fig:eye-b}}
  \hfill
  \subfloat[Double-sided routing]{\includegraphics[height=2.45cm]{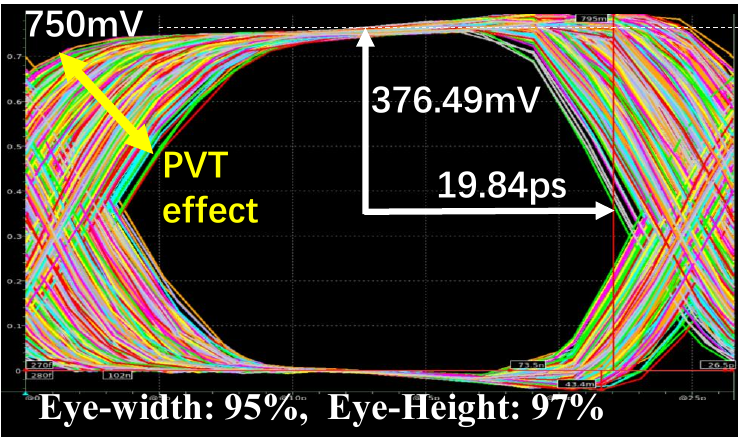}\label{fig:eye-c}}
  \caption{SI mitigation on the worst-case net: eye diagrams with (a) frontside and (b) double-sided routing.}
  \label{fig:eye}
\end{figure}

\subsection{Signal Integrity Analysis}
We count nets with $r_{delay}>0.4$ or $r_{bump}>0.3$ as SI violations. \Cref{tab:ppa} shows that replacing FSPDN with naive BSPDN+FR increases both violations across all benchmarks. On ARM Cortex-A7, the delay- and bump-violation counts increase from 7 to 17 and from 5 to 27, respectively. CSCO reduces the totals across all three benchmarks from 41 to 8 (80.5\%) and from 42 to 5 (88.1\%). On ARM, these counts decrease to 2 and 3, demonstrating that CSCO suppresses aggressor--victim coupling while preserving timing closure.

\Cref{fig:delaybar} shows that the maximum delta delay among the top five critical nets decreases from approximately 30~\si{\pico\second} to below 0.01~\si{\pico\second}. In \Cref{fig:eye-b,fig:eye-c}, CSCO-enabled double-sided routing improves the normalized eye width and height of the worst-case net from 15\%/23\% to 95\%/97\%, corresponding to an absolute eye width of 19.84~\si{\pico\second} and eye height of 376.49~\si{\milli\volt}. These results demonstrate robust high-frequency operation without additional area overhead.

\subsection{Joint BSPDN and Net Allocation}
\begin{figure}[t]
	\centering		
    \subfloat[FSPDN]{\includegraphics[width=0.24\linewidth]{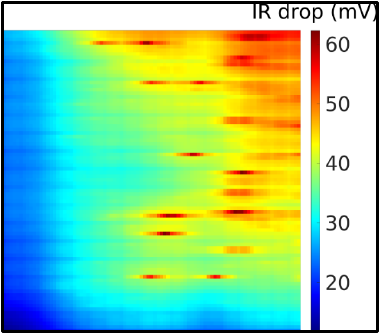} }
    \subfloat[BSPDN-1]{\includegraphics[width=0.24\linewidth]{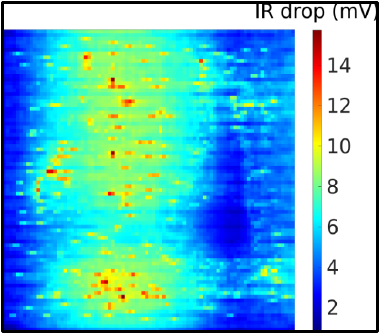} }
    \subfloat[BSPDN-2]{\includegraphics[width=0.24\linewidth]{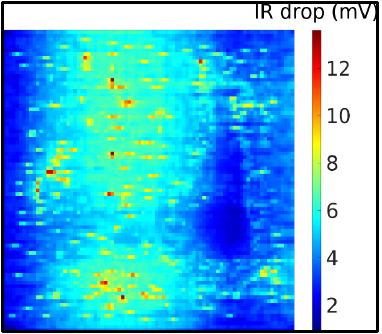} }
    \subfloat[BSPDN-3]{\includegraphics[width=0.24\linewidth]{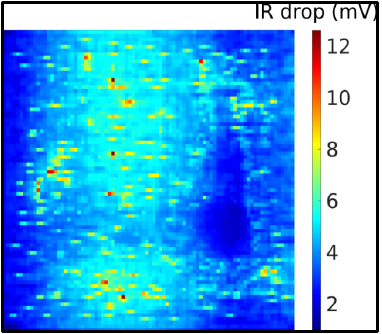} }
	\caption{Static IR-drop maps for SHA256 under FSPDN and three BSPDN configurations (design area: $40 \times 40$~\si{\micro\meter\squared}).}
	\label{fig:pdns}
\end{figure}

CSCO jointly optimizes BSPDN configurations and net allocation. As summarized in \Cref{tab:configs}, progressively wider power stripes reduce line resistance at the cost of higher overall capacitance and less routing space. The static IR-drop maps in \Cref{fig:pdns} illustrate the resulting trade-off between power integrity and routing congestion. Unlike approaches requiring optimization or retraining for each PDN setting, CSCO reuses one surrogate across heterogeneous BSPDN layouts to efficiently explore the joint PDN--routing space and identify Pareto-efficient net-to-layer assignments.

\Cref{tab:tradeoff} shows a non-monotonic trade-off among the configurations. BSPDN-3 minimizes IR-drop but incurs slightly worse timing from higher parasitics; BSPDN-1 achieves the best WNS but higher IR-drop. BSPDN-2 provides the most balanced solution, combining low IR-drop with the best TNS and stable WNS. These results show that CSCO captures configuration-dependent trade-offs and guides topology selection according to design-specific PPA objectives.

\begin{table}[t]  
    \centering
    \caption{BSPDN power-rail configurations. Entries list BSM1/BSM2/BSM3 values; $r$ and $c$ denote resistance and capacitance per unit length.}
    \label{tab:configs}
    \setlength\tabcolsep{5.8pt}
    \fontsize{7}{9}\selectfont 
    \begin{tabular}{c|cccc}
    \toprule
        \textbf{{BSPDN Config}}  & \textbf{{BSPDN-1}} & \textbf{{BSPDN-2}} & \textbf{{BSPDN-3}} \\
        \midrule
        \textbf{Power Rail Width (\si{\micro\meter})}  & 0.10/0.40/1.0 & 0.20/0.60/2.0 & 0.35/0.80/3.0 \\
        \textbf{Power Rail Pitch (\si{\micro\meter})}  & 2.1/1.8/10 & 2.1/1.8/10 & 2.1/1.8/10 \\
        \textbf{Unit $r$ ($\Omega$/\si{\micro\meter})}  & 4.20/0.21/0.016 & 1.86/0.14/0.008 & 1.10/0.10/0.005 \\
        \textbf{Unit $c$ (\si{\femto\farad}/\si{\micro\meter})}  & 0.41/0.23/0.083 & 0.37/0.21/0.153 & 0.39/0.30/0.270 \\
        \bottomrule
    \end{tabular}       
  
\end{table}

\begin{table}[t]
    \centering
    \caption{The solutions given by CSCO on SHA256 with various BSPDN configurations.}
    \setlength\tabcolsep{1.3pt}
    \fontsize{7}{9}\selectfont 
    \resizebox{1.01\columnwidth}{!}{%
    \begin{tabular}{c|c|c|c|c|c|c}
    \toprule
           \textbf{Configs}  &\textbf{IR-Drop (\si{\mV})} &\textbf{\#nTSV}& \textbf{WNS (\si{\nano\second})} & \textbf{TNS (\si{\nano\second})} & \textbf{$r_{delay}>0.4$} & \textbf{$r_{bump}>0.3$} \\
       \midrule
         \textbf{BSPDN-1}&15.76&240& -0.060 & -1.17 & 2 & 1\\ 
         \textbf{BSPDN-2}&14.35&230& -0.061 & -1.09 & 2 & 0\\ 
         \textbf{BSPDN-3}&13.61&210& -0.067 & -1.18 & 1 & 1\\ 
     \bottomrule
    \end{tabular}        
    \label{tab:tradeoff}
    }
\end{table}

\section{Conclusion}

This paper presents CSCO, a BSPDN-aware co-optimization framework that jointly determines PDN topology and dual-side clock/signal routing. By integrating PDN planning with backside-aware net allocation and leveraging data-driven surrogate modeling, CSCO efficiently explores the coupled design space without costly full-routing iterations. Unlike prior methods, CSCO accounts for amplified crosstalk introduced by removing frontside power shields and exploits BSPDN metals to protect SI-critical paths. Evaluations on realistic benchmark designs demonstrate that CSCO achieves significant PPA gains while satisfying power and signal integrity constraints, establishing a unified methodology for next-generation DTCO and 3D ICs.

\clearpage
\bibliographystyle{IEEEtran}
\bibliography{reference}

\end{document}